\documentclass[twocolumn,showpacs,showkeys,floatfix,preprintnumbers,amsmath,amssymb,prl]{revtex4}
\usepackage{graphicx}
\usepackage{amssymb}
\usepackage{multirow}
\usepackage{longtable}

\begin{document}

\title{Highly anisotropic magnetic states of Co dimers
bound to graphene-vacancies}

\author{Hem C. Kandpal}

\author{Klaus Koepernik}

\author{Manuel Richter}

\affiliation{IFW Dresden e.V., PO Box 270116, D-01171 Dresden, Germany}

\date{\today}

\begin{abstract}
The adsorption behavior and the magnetic states of cobalt atoms
and dimers on single vacancies in a graphene sheet are investigated
by means of relativistic density functional calculations.
It is found that local magnetic moments are formed in
both cases, despite strong chemical binding.
Of particular interest are kinetically stable isomers
with two cobalt atoms attached to the same side of the
graphene sheet. Magnetic bi-stability with an anisotropy
barrier of about 50 meV is possible in this geometry.
The feasibility of its preparation is discussed.
\end{abstract}



\maketitle

Nanoscopic magnets with large magnetic anisotropy (MA) are 
of interest both for fundamental 
research~\cite{Gambardella03,Milios07,Balashov09} 
and for technological applications~\cite{Thomson08,Stillrich09}.
Experimental activities to produce such systems were hitherto mainly
focused on one of two known routes, the chemical synthesis of
single-molecule magnets (SMM) or the physical deposition of
magnetic atoms or small clusters on metallic substrates.
Both routes resulted in systems with maximum MA energies (MAE)
somewhat below 10 meV per magnetic atom or per SMM:
Single Co atoms deposited on a Pt(111) surface show an MAE
of 9 meV per Co atom~\cite{Gambardella03};
an anisotropy barrier of 7 meV was reported
for a complex of the Mn$_6$ family of SMM~\cite{Milios07}.

Recently, the MA of four 3d transition metal dimers was
predicted to be up to one order of magnitude larger than these
figures~\cite{Strandberg07,Fritsch08}.
In the case of Co dimers,
adsorption on pristine graphene does obviously not
harm their magnetic properties.
Their MAE has been
estimated by density-functional (DF) calculations~\cite{Xiao09}
to be of the order of 100 meV.
This finding adds a new flavor to the remarkable features
of graphene~\cite{Neto09,Allen10} (G)
which has been praised~\cite{Geim09} for its
electronic, mechanical, optical, and thermal properties.

Its structural robustness~\cite{Geim09}, impermeability~\cite{Bunch08},
chemical tolerance~\cite{Shen09},
and reassemble properties~\cite{Paek09}
distinguish G as a salient substrate material.
As such, it has been applied to adsorb metal atoms~\cite{Gierz08,Hwang09}
and clusters~\cite{Muszynski08}.
Several theoretical studies confirmed that magnetic moments
are to be expected
on transition metal atoms and dimers including
cobalt adsorbed on G~\cite{Chan08,Johll09,Cao10,Zhang12}.
Thus, Co$_2$-G seems to be suited for ultrahigh-density
magnetic recording~\cite{Xiao09}.

Unfortunately, the binding of transition metal atoms and dimers
to pristine G is relatively weak. In the case of Co
dimers, the DF binding energy amounts to about 0.7 eV.~\cite{Xiao09}
However, a yet smaller activation energy of 0.24 eV (upper limit)
for surface migration
was measured for gold atoms on graphite~\cite{Anton98}
and migration barriers in the range
of 0.01 to 0.8 eV were evaluated~\cite{Chan08,Krasheninnikov09} by DF 
calculations for transition metal adatoms on G.
At room temperature, Co atoms are hence expected to float
on the carbon layer~\cite{Krasheninnikov09}.
Indeed, STM images of Co atoms on pristine G were
only obtained at 4.2 Kelvin~\cite{Brar10}, hitherto.

Defect structures of carbon are highly reactive.
Thus, one possibility to bind transition metal atoms or dimers
more strongly consists in a deliberate creation of structural
defects, e.g. single vacancies with dangling bonds,
in the G sheet before the adsorbate is deposited.
An established experimental route
to produce isolated vacancies in graphene-like systems
is irradiation with low-energy Ar$^+$ ions~\cite{Lee99}.
After annealing, most defects produced in this way
were found to be single vacancies~\cite{Ugeda10}.

For the case of Co atoms, a binding energy to a single 
vacancy in graphene (SVG) of about 8 eV
was predicted~\cite{Krasheninnikov09}, by far large
enough to ensure stability at room temperature.
Strong chemical binding, on the other hand, often
spoils magnetism. In particular, orbital magnetism which
is chiefly responsible for the MA is known to be quenched
by hybridization and crystal fields.
While the existence of spin magnetism was 
predicted for atomic impurities of several transition metals
in G including Co-SVG~\cite{Krasheninnikov09,Santos10},
we are not aware of related investigations concerning
orbital magnetic properties.
It is thus both demanding and interesting to find
magnetic graphene-based
nanostructures that are structurally stable on the one hand
and show a large orbital moment or even a large MA on the other hand.

Here, we present Co$_2$-SVG as an example of a nanostructure
that possibly fulfills both contradicting requirements.
The related system Co-SVG is demonstrated to be a
counterexample where the strong binding suppresses the MA.
As a specific intricacy, the desired large MA of
Co$_2$-SVG is not present in the ground state but in kinetically stable
isomers which can be prepared in the most simple way
by atom-wise deposition.
The suggested preparation can be considered as a 
combination of the two known routes to produce nanomagnets with large MA,
the chemical (SMM) and the physical (adsorbate deposition) one.

The DF calculations were performed with the
all-electron full-potential local-orbital (FPLO) code~\cite{Koepernik99},
version 9.00-34.
The exchange-correlation energy functional was evaluated
using the standard parameterization~\cite{Perdew96}
of the generalized gradient approximation (GGA).
Orbital magnetic properties were calculated in a four-component
fully relativistic mode.
To obtain an upper estimate for
the MAE~\cite{Xiao09}, orbital polarization
corrections~\cite{Eriksson90} (OPC) were applied in some of the calculations.
The technical parameters of the calculations are described in the
Online Supplement~\cite{Supplement}.

\begin{figure}[tb!]
\includegraphics[width=0.7\columnwidth]{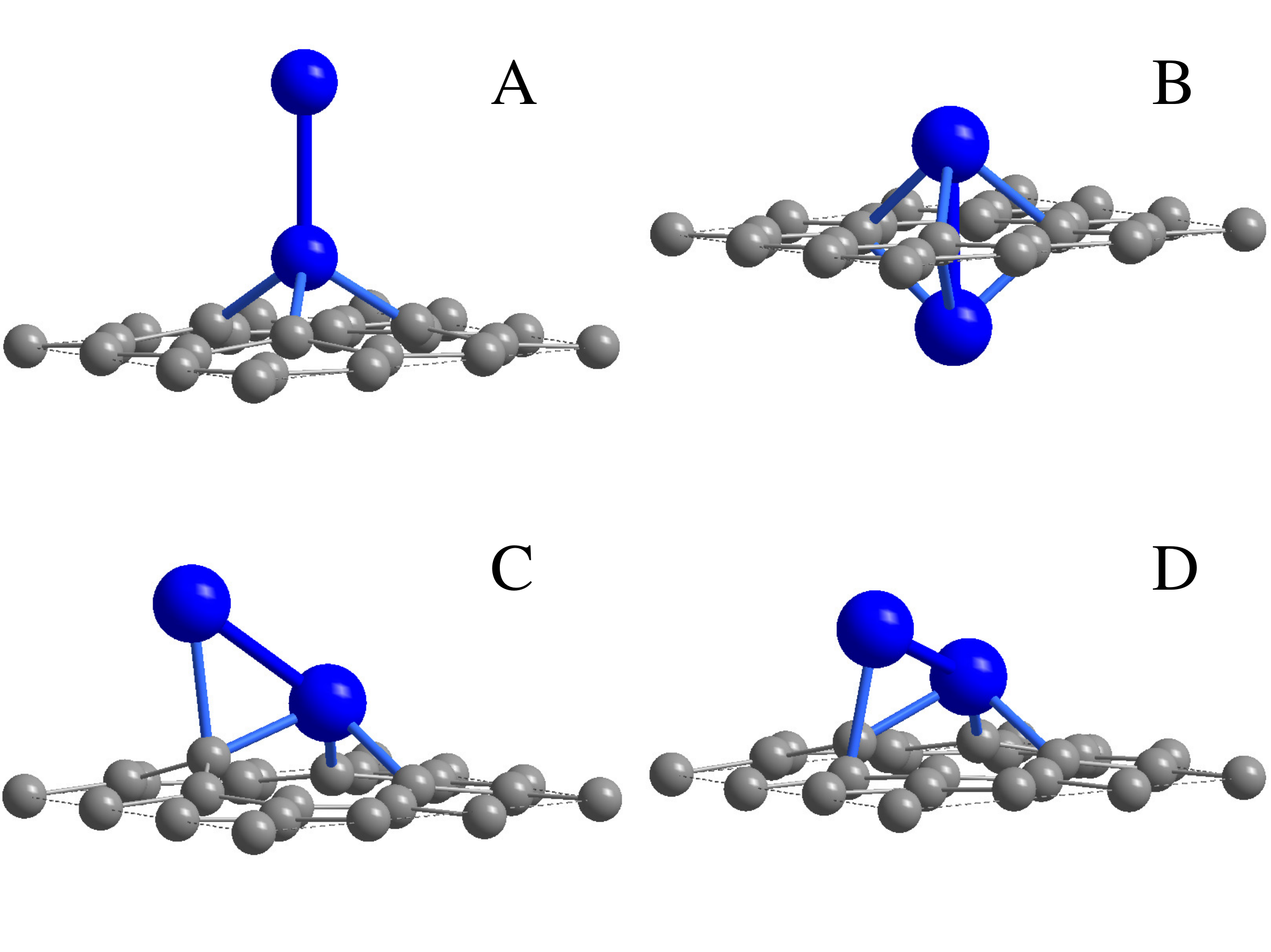}
\caption{(Color online)
Four isomers (A)-(D) of Co$_2$-SVG obtained by relaxation of five initial
geometries~\cite{Supplement}.
Large blue spheres: cobalt atoms; small gray spheres: carbon atoms.}
\label{fig:str1}
\end{figure}

Recent DF results indicate
that {\em single} Co atoms on {\em pristine} G
are magnetically bistable with an MAE-barrier
in the 10-meV range~\cite{Xiao09}.
Further, using GGA it has already been found that Co atoms bound to an
SVG keep a spin moment of 1~$\mu_{\rm B}$, while
Fe and Ni atoms become non-magnetic~\cite{Krasheninnikov09,Santos10}.
These predictions tempted us to check the magnetic anisotropy
of Co-SVG, before considering the more complex case of Co dimers.
The values of spin moment, $\mu_s$(Co-SVG) = 0.92 $\mu_{\rm B}$,
energy needed for Co detachment,
$E_{\rm Co-det}$(Co-SVG) = 8.0 eV, and Co-C bond
length, $d^{\rm Co-C}$(Co-SVG) = 1.76 \AA{}, found in our 
calculations~\cite{Supplement} confirm the available
literature data~\cite{Krasheninnikov09,Santos10,Boukhvalov09}.
By checking the orbital magnetic properties,
we however find negligible magnitudes of the orbital
moment, $\mu_l$(Co-SVG) $<$ 0.01 $\mu_{\rm B}$, and of the MAE 
($<$ 0.1 meV) in both GGA and GGA+OPC approaches.
Thus, we conclude that the orbital magnetism of {\em single}
Co atoms is quenched by binding to an SVG.

\begin{table}
\begin{tabular}{l | c | c | c | c | c | c }
Co$_2$-SVG &(A0) & (AI) & (AII) & (B) & (C) & (D) \\
\hline
$E_{\rm tot}$ (eV) & 0.9 & 1.0 & 1.2 & 0 & 1.1 & 1.4\\
$\mu_s$ ($\mu_{\rm B}$) & 2.0 & 2.0 & 2.0 & 1.7 & 2.0 & 1.8\\
$E_{\rm spin}$ (eV) & 0.7 & 0.8 & 1.0 & 0.1 & 0.5 & 0.1 \\
\end{tabular}
\caption{Relative total energies, $E_{\rm tot}$, and total 
spin moments, $\mu_s$, of the structural
isomers of Co$_2$-SVG, Fig.~\ref{fig:str1}.
$E_{\rm spin}$ denotes the stabilization
energy of spin magnetism, i.e., the energy difference between
non-magnetic and magnetic state. Structure (A) can be stabilized in
three different electronic states: (A0), (AI), and (AII).}
\label{tab:table1}
\end{table}

There are several structures for the interaction of {\em two}
Co atoms with SVG imaginable. We considered five initial 
geometries~\cite{Supplement}.
Relaxation of these geometries resulted in four (meta)stable
structures, (A) to (D) in  Fig.~\ref{fig:str1}.
In the case of (A), we found three different electronic isomers
(A0), (AI), and (AII). They are distinguished by specific 3d-orbital
occupations at the ``remote'' Co atom (r-Co) which is not directly
bound to the vacancy.
The relative total energies and the spin moments of all isomers
are given in Tab.~\ref{tab:table1}, their structural details are compiled 
in the Online Supplement~\cite{Supplement}.

All identified Co$_2$-SVG isomers are magnetic with spin $S=1$.
This is a notable difference to the $S=2$ states of free Co
dimers~\cite{Strandberg07} and of Co$_2$-G~\cite{Xiao09}.
The lower spin state of Co$_2$-SVG mirrors the stronger chemical
bonding of Co at the vacancy compared to the pristine carbon hexagon.

\begin{table}
\begin{tabular}{l | c | c | c | c | c }
   $m$ & -2 & -1 & 0 & 1 & 2 \\
\hline
(A0) & 0.0 & 0.9 & 0.1 & 0.9 & 1.0    \\
(AI) & 0.5 & 0.5 & 0.8 & 0.5 & 0.5    \\
(AII)& 0.5 & 0.9 & 0.1 & 0.9 & 0.5    \\
(C)  & 0.6 & 0.5 & 0.8 & 0.4 & 0.6    \\
\end{tabular}
\caption{Orbital-resolved occupation numbers of 3d spin-down states 
of r-Co for the three electronic isomers of the geometry (A)
and for geometry (C).
The quantum numbers $m$ refer to real spherical harmonics.
}
\label{tab:orb}
\end{table}

Table~\ref{tab:orb} compiles the 3d spin-down orbital occupation numbers
of r-Co for (A0-II). In all three isomers, the spin moment resides on r-Co
and the Co atom at the vacancy (v-Co) is nearly spin-compensated.
Thus, the spin-up channel of r-Co is almost completely occupied.
State (A0) is lower in energy than (AI) and (AII), but its charge density
lacks axial symmetry. Using a GGA+$U$ mode with $U,\; J \rightarrow 0$ we
were able to stabilize all three states in a scalar relativistic GGA
calculation for a free Co atom.
The energy sequence is now (A0, free) - (AII, free) at 0.1 eV - (AI, free)
at 0.2 eV. The correct atomic ground state has an axially symmetric
charge density like (AI, free) or (AII, free), but unlike (A0, free).
If spin-orbit interaction (s-o) is switched on, we find $L=0$ for
(A0, free), $L=3$ for (AI, free), and $L=2$ for (AII, free).
Considering Hund's second rule, the best GGA approximation to 
the atomic ground state is (AI, free), the second best is (AII, free).
Both facts, its broken axial symmetry and its incompatibility with
Hund's second rule, lead us to disregared the state (A0, free) as unphysical.
Its low energy is due to self-interaction of the non-spherical part of the
charge density, as discussed earlier by Brooks {\em et al.}~\cite{Brooks97}.
Bonding of the free Co atom to Co-SVG slightly shifts the energies of the
three electronic isomers, but hardly changes their orbital occupations.
We therefore exclude (A0) from the further discussion.

Structure (B), where the two Co atoms are located on opposite
sides of the G plane, is the ground-state isomer.
However, we will now
provide arguments that it is technically easier to produce a structure
with both Co atoms on the same side of the carbon
layer and that this situation is kinetically stable.

Deposition of small Co amounts on a graphene-like
surface with aforehand-prepared single vacancies
will yield Co-SVG under ambient conditions, since Co atoms
are mobile on pristine surface areas.
A fraction of the Co atoms may agglomerate
[Co-G $\rightarrow$ G + hcp-Co],
but the related energy gain per Co atom~\cite{Supplement},
$E_{\rm Co-agg} = - 4.4$ eV, has a
smaller absolute value than the energy gained by binding to the SVG
[Co-G + SVG $\rightarrow$ G + Co-SVG],
$E_{\rm b}({\rm Co-SVG}) = - 6.8$ eV.
We further note, that Co-SVG is expected to be stable against
migration at room temperature. Related activation energies
for Pt-SVG and Au-SVG were found to be close to 2.5 eV by high-temperature
transmission electron microscopy~\cite{Gan08}.

\begin{figure}
\includegraphics[width=0.9\columnwidth]{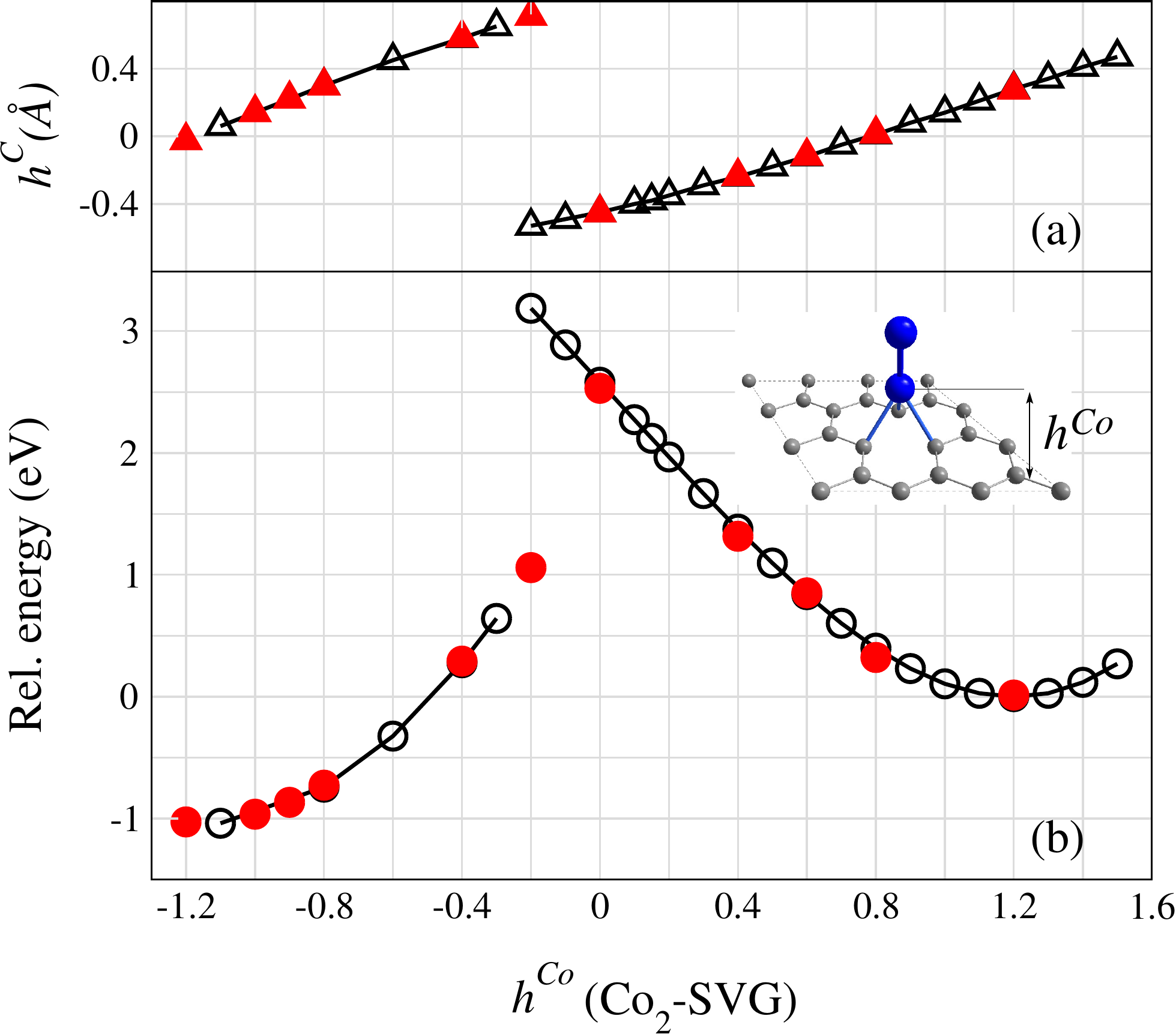}
\caption{(Color online)
Diffusion barrier between Co$_2$-SVG(AI) and Co$_2$-SVG(B).
The inset shows the considered structure.
The height of the Co atom next to the vacancy, $h^{\rm Co}$, was fixed
(abscissa of the figure).
Black (open) symbols: space group 156 (P3m1),
red (full) symbols: space group 1 (P1). 
Relaxation was allowed for the height of the three neighboring
carbon atoms, $h^{\rm C}$ of Panel (a), their lateral position,
and the distance between the two Co atoms (not shown).
Panel (b) displays the total energy relative to the energy of Co$_2$-SVG(AI).
The two minima in the energy curves are related to the two isomers
(AI) and (B), the diffusion barrier amounts to $2\ldots2.5$ eV,
estimated by extrapolation of the left branch.  }
\label{fig:diffusion}
\end{figure}

If Co-deposition is continued after saturation of the vacancies,
a large fraction of the mobile Co atoms will attach to Co-SVG
and form Co$_2$-SVG(AI) or one of its quasi-degenerate isomers
(AII) or (C). The reaction [Co-G + Co-SVG $\rightarrow$
G + Co$_2$-SVG(AI)] gains an energy~\cite{Supplement}
of $E_{\rm b}({\rm Co_2-SVG}) = - 1.2$ eV.
Agglomeration of the Co atoms with a gain of -4.4 eV
would be energetically yet more favorable, but obviously
the formation of Co$_2$-SVG is kinetically more likely at low 
deposition rates and sufficient supply of Co-SVG sites.
The substrate temperature should be low enough to avoid detachment
of Co from Co$_2$-SVG (below 300 K, estimated from 
$|E_{\rm b}({\rm Co_2-SVG})| = 1.2$ eV),
but high enough to ensure mobility of the deposited Co atoms on G.

Finally, we check if Co$_2$-SVG(AI) is kinetically stable against
diffusion of the Co atom next to the vacancy to the other side of the
G sheet [Co$_2$-SVG(AI) $\rightarrow$ Co$_2$-SVG(B)].
Fig.~\ref{fig:diffusion} presents the geometry and the energy of
structures along the diffusion path.
We find an energy barrier of more than 2 eV, large enough to ensure
stability beyond detachment of the upper Co atom.
This large barrier arises from the fact that the Co atomic radius is 
much larger than the radius of the missing carbon atom.
Thus, Pauli repulsion on the one hand and stiffness of the carbon plane
on the other hand hinder the diffusion.
{\em We conclude that preparation of Co$_2$-SVG with both Co atoms
located at the same side of the C plane is possible and that this
structure is stable up to about 300 K.}

Considering the size of the discussed self-energy error for
systems containing almost free atoms, we are
not able to judge if (AI), (AII), or (C) is the ground state isomer.
We however tend to exclude geometry (D), which lies $0.4$ eV above (AI),
from the further discussion.
We also note that the magnetic state of (AI), (AII),
and (C) is more stable than that of the other isomers, cf. the spin
stabilization energies in Tab.~\ref{tab:table1}.

\begin{table}
\begin{tabular}{c | c | c | c | c }
orientation &        & (AI) & (AII) & (C) \\
\hline
001 & $\mu_s$(v-Co)  & 0.0  & -0.1 (-0.1) & 0.0 (0.0) \\
100 & $\mu_s$(v-Co)  & 0.0  & -0.1 (-0.1) & 0.0 (0.0) \\
\hline
001 & $\mu_s$(r-Co)  & 2.1  & 2.3 (2.3)   & 2.0 (2.0) \\
100 & $\mu_s$(r-Co)  & 2.1  & 2.3 (2.3)   & 2.0 (2.0) \\
\hline
001 & $\mu_l$(r-Co)  & 0.3  & 2.0 (2.1)   & 0.2 (2.2) \\
100 & $\mu_l$(r-Co)  & 0.2  & 0.2 (0.8)   & 0.2 (0.6) \\
\hline
    & MAE (meV)      & 0.1  & 63 (53)   & 0.6 (44) \\
\end{tabular}
\caption{
Results from fully relativistic GGA calculations for Co$_2$-SVG(A)
and Co$_2$-SVG(C).
The notations 001 and 100 refer to magnetic moment orientations 
perpendicular and parallel to the C plane, respectively.
Magnetic spin moments $\mu_s$ and orbital moments $\mu_l$ are
projected on the cobalt atoms next to the vacancy (v-Co) and on the
other Co atom (r-Co) and given in $\mu_{\rm B}$.
The MAE is the difference between total energies
obtained for the two chosen moment orientations, MAE = $E(100) -E(001)$,
and refers to the whole structural unit with two Co atoms.
The in-plane anisotropy is small and can be neglected.
Results in parentheses refer to GGA+OPC calculations.}
\label{tab:table3}
\end{table}

We now turn the attention to orbital magnetic properties.
Tab.~\ref{tab:table3} compiles results for the kinetically stable
low-lying isomers.
In the case of (AI), any self-consistent GGA+OPC calculation
converged to the state (AII).
This might be an indication that (AI) is not realized in nature.
Most remarkably, the MAE of (AII) and of (C) in the GGA+OPC
approach is predicted to be similarly
high as that of Co$_2$-G~\cite{Xiao09}, though in the present case
the Co dimer is much more tightly bound to the substrate and thus has a
much better chance to be accessible to experimental realization.

Strong chemical bonding claims its price, though.
The atomic-site projected magnetic moments of the Co atom next to
the vacancy are almost completely quenched,
Tab.~\ref{tab:table3}.
However, large spin moments ($S=1$) persist on r-Co
due to the low coordination of this atom.
For (AII) and for (C) in OPC, also the orbital moment is large, $L=2$.
Magnetic moment orientation perpendicular to the easy (001) axis
quenches the orbital moment but requires a large energy, the MAE,
which is related to the orbital moment reduction~\cite{Bruno89,Xiao10}.

Among the two Co atoms, a distribution of tasks takes place:
v-Co atom is strongly bound to SVG and serves as an anchor for r-Co;
the remote r-Co, on the other hand, preserves atomic-like properties
and carries those electronic states which produce an ultimately
high MAE. From this distribution of tasks it is clear that atoms
of elements other than cobalt could be used as anchor atoms and/or
remote atoms. Thus,
Co$_2$-SVG is most likely only one example of a whole class of
unknown highly anisotropic nanomagnets.

To summarize, we suggest that deposition of Co atoms on
SVG can be used to prepare a
structure with two Co atoms attached to each vacancy.
We demonstrated its kinetic stability and found strong
indications that it will show an exceptionally high MA.
Single Co atoms on Pt also show large 
MA~\cite{Gambardella03}, but their technological use might
be restricted by a high spin tunneling probability~\cite{Balashov09}.
For the present system, the
hybridization of the magnetically active atom with its surrounding
is much smaller
than in the case of a metallic substrate, since the chemical bonding
is mediated by a single anchor atom.
Thus, Co$_2$-SVG might even be of technological interest for magnetic
data storage.

We finally note that Co$_2$-SVG should be considered
as one didactic example of a whole class of possible structures,
consisting of an organic substrate with dangling bonds and a few
metal atoms.
The ``remote'' atoms show
strong orbital magnetism, while the ``anchor'' atoms mediate the bonding
to the substrate.
The production of such systems could make use of a
self-assembling step: the saturation of the dangling bonds by anchor atoms.

Financially supported by DFG grant RI932/6-1.

\bibliography{references}

\clearpage

\setcounter{secnumdepth}{3}

\setcounter{table}{0}
\setcounter{figure}{0}

\renewcommand{\thetable}{S\Roman{table}}
\renewcommand{\thefigure}{S\arabic{figure}}
\renewcommand{\thesection}{}
\renewcommand{\thesubsection}{S\arabic{subsection}}

\section*{Supplementary material}

This Supplement reports technical details of the calculations,
results of the optimization
of geometry and/or spin state, definitions of reaction energies
and convergence checks for the simulation cell size.

\subsection*{Technical details of the calculations}

The convergence with respect to the $k$-space grid density
was carefully checked. 
Using a linear tetrahedron method with Bl\"ochl corrections,
the final calculations were performed with
$9\times 9\times 1$ $k$-points in the full Brillouin zone
(geometry optimization) or with $30\times 30\times 1$ $k$-points
(orbital magnetism).
The basis set comprised Co (3s, 3p, 3d, 4s, 4p, 4d, 5s) and
C (1s, 2s, 2p, 3s, 3p, 3d) states.
The modified G sheet was modeled with a $3\times 3$ supercell of
fixed edge length 7.392~\AA{}
in the $x$-$y$ plane and with 12~\AA{} spacing along the $z$ direction.
Calculations with a $4\times 4$ cell and with
16~\AA{} spacing did not show any significant difference, see below.

Geometry optimization was carried out using forces evaluated in
a scalar relativistic mode with a convergence criterion of 1 meV/\AA{}.
Fixed spin-moment calculations were employed to exclude the existence
of spin states not found by the unrestricted self-consistent calculations.

\subsection*{Optimization of geometry and/or spin state}

\begin{figure}
\includegraphics[width=0.15\textwidth]{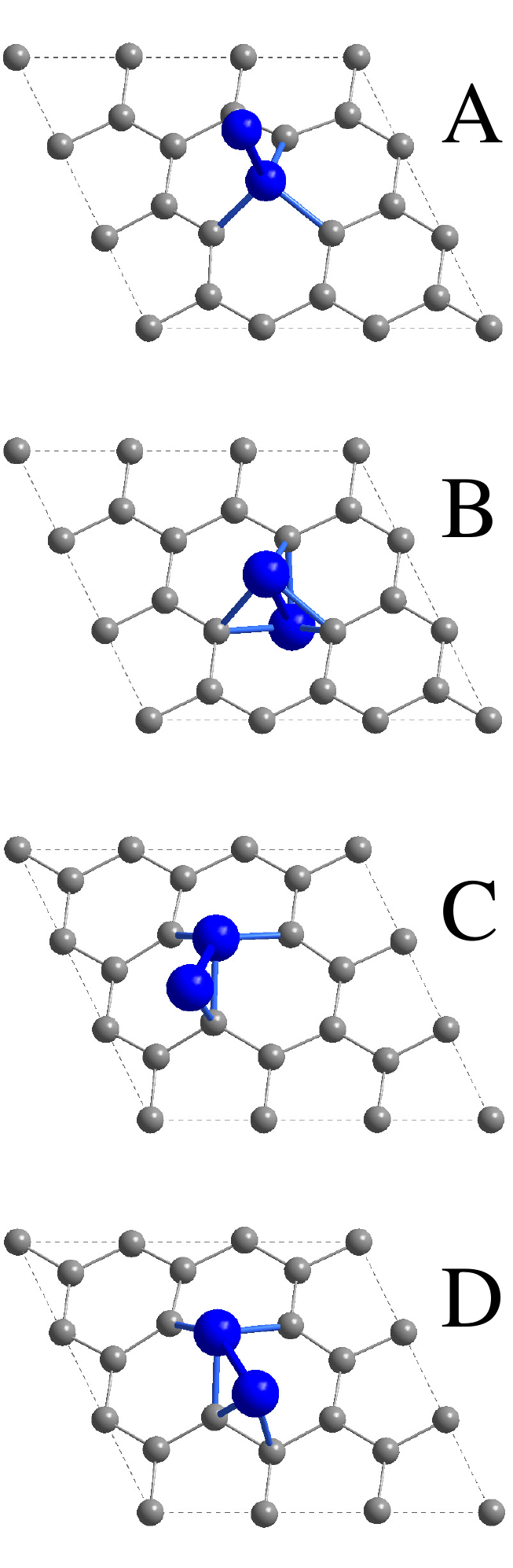}
\caption{
Four isomers (A)-(D) of Co$_2$-SVG obtained by relaxation of five initial
geometries, see \protect Fig.~\ref{str3}.
Blue spheres denote cobalt atoms and gray spheres denote carbon atoms.
Fig.~\ref{fig:str1} of the main text shows the same structures in
a different view.}
\label{fig:str2}
\end{figure}

The geometry and/or the spin magnetic state of eight different
systems were optimized with respect to the total GGA
energy: a free Co atom, a free Co dimer (Co$_2$), bulk
Co metal in the hexagonal close-packed structure (Co-hcp),
a Co atom on pristine graphene (Co-G),
a Co dimer on pristine graphene (Co$_2$-G),
graphene with a single vacancy (SVG),
Co-SVG, and Co$_2$-SVG with four identified structural isomers,
called (A, B, C, D), see Fig.~\ref{fig:str2}
and Fig.~\ref{fig:str1} of the main text.
In the isomer Co$_2$-SVG (B), the two Co atoms are
located on opposite sides of the graphene sheet. Thus, this
isomer is also referred to as Co-SVG-Co.

All graphene-derived systems were treated in a $3\times 3$
structural unit cell.
For comparison, the total energy of pristine graphene (G) is computed in
a fixed geometry, according to the fixed unit cell size used in
the other calculations.
All energies and spin moments in Tabs.~\ref{SI} and \ref{SII}
refer to the considered structural unit,
i.e., to 18 carbon atoms for G, to 18 carbon atoms and 
1 (2) cobalt atom(s) for Co-G (Co$_2$-G), to 17 carbon atoms for SVG, and
to 17 carbon atoms and 1 (2) cobalt atom(s) for Co-SVG (Co$_2$-SVG).

All calculations were performed with the code and settings defined in the
main text, exceptions are mentioned in the column `remarks'.
For Co atoms and dimers, the molecular mode with free boundary
conditions was applied. All other calculations were performed
with three-dimensional periodic boundary conditions.

\begin{figure}
\includegraphics[width=0.15\textwidth]{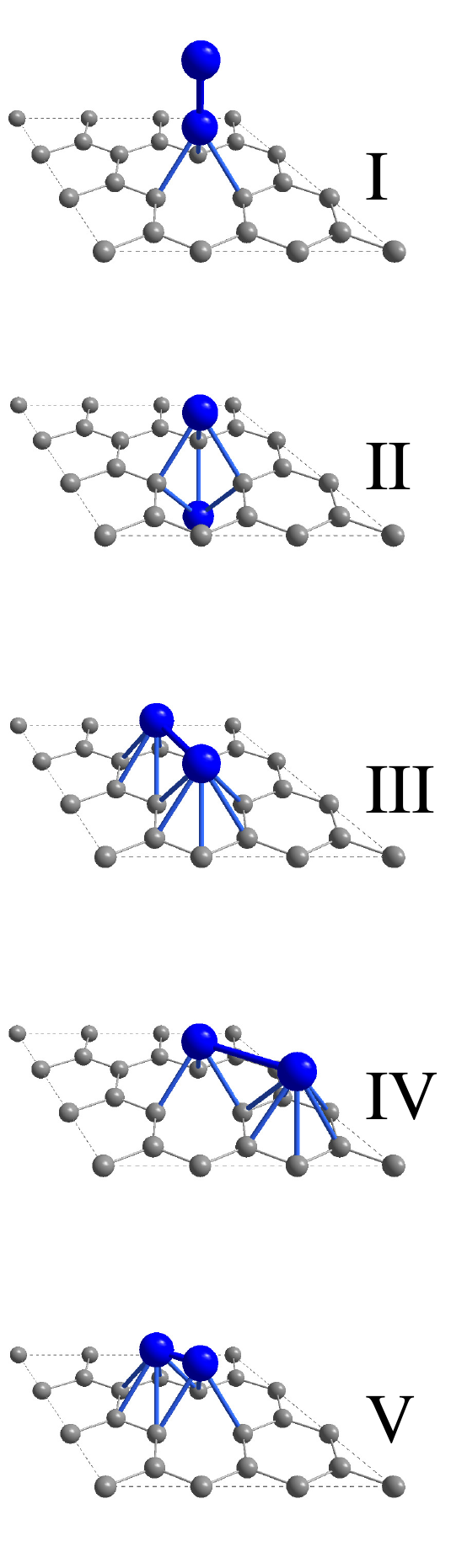}
\caption{
Five chosen initial geometries (I)-(V) of Co$_2$-SVG.}
\label{str3}
\end{figure}

In the atomic calculations, the 3d and 4s states turn out to be degenerate. 
The lowest GGA energy corresponds to a configuration 3d$^{7.6}$4s$^{1.4}$.
Spin-orbit coupling has a stronger influence on the atomic total energies
than on the total energies of atoms in a cluster, molecule, or bulk compound.
We included related data obtained in fully relativistic mode
for the Co atom and the Co
dimer in Tab.~\ref{SI} but used the values obtained in scalar relativistic
mode to calculate binding energies for the sake
of comparability with other published (scalar relativistic)
DF energies~\cite{Krasheninnikov09}.
Binding energies related to atomic Co would have to be corrected by
the amount $E({\rm Co-atom}) - E_r({\rm Co-atom})$ = 0.08 eV.
The correction related to a Co dimer amounts to less than 0.02 eV, see
Tab.~\ref{SI}.

Fig.~\ref{str3} shows the five chosen initial structures
used to identify isomers of Co$_2$-SVG by structural relaxation.
The initial SVG structure
was always prepared by removing one carbon atom from the pristine
graphene sheet without relaxation.
The two Co atoms were initially placed in the following positions to form the 
initial structures (I) to (V): \\
(I) both above the vacancy site at heights of 1.5 and 3.5 \AA{}; \\
(II) above and below the vacancy site at heights of 1.5 and -1.5 \AA{}; \\
(III) above the centers of two pentagons next to the vacancy
at the same height of 1.5 \AA{}; \\
(IV) above the vacancy site and above the center of a neighboring hexagon
at the same height of 1.5 \AA{}; \\
(V) above the vacancy site and above the center of a pentagon
next to the vacancy at the same height of 1.5 \AA{}.

The initial geometries for Co-G and Co-SVG were prepared by positioning
the cobalt atom above the center of a hexagon at a height of 1.5 \AA{}
and above the vacancy site at a height of 1.5 \AA{}, respectively.
In the case of Co$_2$-G, the Co atoms were placed above the center of 
a hexagon at heights of 1.5 and 3.5 \AA{}.

A summary of results for all optimized geometries is given
in Tabs.~\ref{SI} and ~\ref{SII}.
The calculated atomic distances $d$ and heights $h$ in
Tabs.~\ref{SI} and \ref{SII} refer to nearest neighbor
distances and to the distances to the $x$-$y$ plane, respectively.
In particular, $d^{\rm C-C}$ denote the distances between pairs
of the three carbon atoms next to the vacancy, $h^{\rm C}$ is the displacement
of these atoms from the $x$-$y$ plane,
$d^{\rm Co-C}$ are the distances between the cobalt atom next to the 
vacancy and the related carbon atoms, and
$h^{\rm Co}$ is the distance of the cobalt atom next to the
vacancy from the $x$-$y$ plane.

A notable structural deformation was found at the vacancy site
for the Co$_2$-SVG-isomers (A), (C), and (D), where
the three carbon atoms next
to the vacancy are pulled out up to 0.6 \AA{} from the graphene plane.

In the case of structure (A) we found three electronic isomers.
Isomer (A0) was obtained if the calculation was performed in space group 1,
the other two, (AI) and (AII), in space group 156, using different 
starting conditions.

\subsection*{Reaction energies}

We define reaction energies of the following processes, referred to
in the main text:\\[3mm]
Energy needed for the detachment of Co atoms from SVG
(reaction Co-SVG $\rightarrow$ Co-atom + SVG),
\begin{eqnarray}
E_{\rm Co-det}{\rm (Co-SVG)} & = &  \nonumber \\
E{\rm (Co-atom)} + E{\rm (SVG)} - & & \nonumber \\
- E{\rm (Co-SVG)} & = & 7.96 \; {\rm eV} \; .
\end{eqnarray}

\noindent
Energy gained by agglomeration of Co atoms on graphene 
(reaction Co-G $\rightarrow$ G + hcp-Co),
\begin{eqnarray}
E_{\rm Co-agg} & = & \nonumber \\
E{\rm (G)} + E{\rm (Co-hcp)} - & & \nonumber \\
- E{\rm (Co-G)} & = & - 4.43  \; {\rm eV} \; .
\end{eqnarray}

\noindent
Energy gained by binding of Co atoms on graphene to SVG
(reaction Co-G + SVG $\rightarrow$ G + Co-SVG),
\begin{eqnarray}
E_{\rm b}({\rm Co-SVG}) & = &  \nonumber \\
E{\rm (G)} + E{\rm (Co-SVG)} - & & \nonumber \\
- E{\rm (Co-G)} - E{\rm (SVG)} & = & - 6.76 \; {\rm eV} \; .
\end{eqnarray}

\noindent
Energy gained by binding of Co atoms on graphene to Co-SVG.
For reasons explained in the main text, we take isomer (AI)
as reference
(reaction Co-G + Co-SVG $\rightarrow$ G + Co$_2$-SVG(AI)),
\begin{eqnarray}
E_{\rm b}({\rm Co_2-SVG}) & = &  \nonumber \\
E{\rm (G)} + E{\rm (Co_2-SVG(AI))} - & & \nonumber \\
- E{\rm (Co-G)} - E{\rm (Co-SVG)} & = & - 1.15 \; {\rm eV} \; .
\end{eqnarray}

\subsection*{Convergence of the simulation cell}

We checked the magnetic moment and binding energy of Co-SVG with 
a $4\times 4$ supercell and found that the ground state 
has a magnetic moment of 0.99~$\mu_{\rm B}$. 
The Co-detachment energy of this structure amounts to
7.99~eV. These values are almost the same as those obtained with
a $3\times 3$ supercell, 0.92~$\mu_{\rm B}$ and 7.96~eV.

In the case of SVG, the $3\times 3$ supercell yields equal distances
between the carbon atoms next to the vacancy and a vacancy formation
energy,
\begin{eqnarray}
E_{\rm vac} (3\times 3) & = & \nonumber \\
E{\rm (SVG,} 3\times 3) - E{\rm (G,} 3\times 3) 
\cdot 17/18 & = & 7.95 \; {\rm eV} \; .
\end{eqnarray}
If a $4\times 4$ supercell is used, the so-called 5-9 structure
is obtained where two of the carbon atoms next to the vacancy 
form a loose pair. Accordingly, the vacancy formation energy
is about $0.1$ eV smaller than in the former case,
\begin{eqnarray}
E_{\rm vac} (4\times 4) & = & \nonumber \\
E{\rm (SVG,} 4\times 4) - E{\rm (G,} 4\times 4) 
\cdot 31/32 & = & 7.84 \; {\rm eV} \; .
\end{eqnarray}

The dependence of $E{\rm (SVG)}$ on the size of the simulation cell
is compensated by an opposite dependence of $E{\rm (Co-SVG)}$.
Thus, $E_{\rm Co-det}{\rm (Co-SVG)}$ is almost independent of the
cell size.
We conclude that the reaction energies are
reliable at a level of 0.1 eV, if $3\times 3$ supercells
are used.

\clearpage

{\small
\begin{table}
\begin{tabular}{|l|l|l|l|l|l|}
\hline
system & space       & geometry             & total        & spin   & remarks   \\
       &       group &          parameters  &       energy & moment &           \\
       &             &       [\AA{}]        &  [eV]        &  [$\mu_{\rm B}$] & \\
\hline
Co atom &   &  & $E$ = -37916.34 & 3.00 &  scalar relativistic \\
Co atom &   &  & $E_{\rm r}$ = -37916.42 & 3.00 &  fully relativistic \\
\hline
Co$_2$  &   & $d^{\rm Co-Co}$ = 1.989 & $E$ = -75836.44 & 4.03 & scalar relativistic \\
Co$_2$  &   & $d^{\rm Co-Co}_{\rm r}$ = 1.989 & $E_{\rm r}$ = -75836.45 & 4.03 & fully relativistic, \\
        &   &                                 &                         &      & distance fixed \\
\hline
Co-hcp  & 194           & $a$ = 2.5071  & $E$ = -37921.97 & 1.62 & experimental geometry \\
        & (P6$_3$/mmc)     & $c$ = 4.0695  & 		 &      & $k$-mesh $96\times 96\times 60$ \\
	&                  &               &             &      & scalar relativistic \\
\hline
G  & 156 (P3m1) & $d^{\rm C-C}_{\rm G}$ = 1.4226 & $E$ = -18663.46   & 0 & experimental geometry \\
   &              &                                &                  &   & $k$-mesh $60\times 60\times 1$ \\
   &              &               &                &      	      & scalar relativistic \\
\hline
\end{tabular}
\caption{Compilation of reference energies and other properties of Co atoms,
Co dimers, bulk Co, and pristine graphene.}
\label{SI}
\end{table}

\begin{table}
\begin{tabular}{|l|l|l|l|l|}
\hline
system & space group & geometry parameters & total energy & spin \\
       &             &      [\AA{}]        &  [eV]        &  moment   \\
       &             &                     &              &  [$\mu_{\rm B}$]  \\
\hline
Co-G   & 183 (P6mm) & $d^{\rm Co-C}$ = 2.111 & $E$ = -56581.00 & 1.06  \\
        &   & $h^{\rm Co}$ = 1.560 &             &  \\
\hline
Co$_2$-G       & 183 (P6mm) & $d^{\rm Co-Co}$ = 2.054 & $E$ = -94500.57 & 4.06 \\
               &   & $d^{\rm Co-C}$ = 2.250 & &  \\
               &   & $h^{\rm Co}$ = 1.743 & &  \\
\hline
SVG    & 1 (P1) & $d^{\rm C-C}$ = 2.545, 2.546, 2.545 & $E$ = -17618.65 & 1.04  \\
       &   & $h^{\rm C}$ = 0 &  &    \\
\hline
Co-SVG  & 1 (P1) & $d^{\rm Co-C}$ = 1.762, 1.762, 1.762 & $E$ = -55542.96 & 0.92  \\
        &   & $h^{\rm Co}$ = 1.144 & &  \\
\hline
Co$_2$-SVG     & 1 (P1) & $d^{\rm Co-Co}$ = 2.250 & $E$ = -93461.76 & 2.00 \\
(A0)           &   & $d^{\rm Co-C}$ = 1.763, 1.763, 1.763 & &  \\
               &   & $h^{\rm Co}$ = 1.195 & &  \\
               &   & $h^{\rm C}$ = 0.270, 0.275, 0.275 & &  \\
\hline
Co$_2$-SVG     & 156(P3m1) & $d^{\rm Co-Co}$ = 2.280 & $E$ = -93461.65 & 2.00 \\
(AI)           &           & $d^{\rm Co-C}$ = 1.764        & &      \\
	       &   & $h^{\rm Co}$ = 1.200                  & &      \\
	       &   & $h^{\rm C}$ = 0.275                   & &      \\
\hline
Co$_2$-SVG     & 156(P3m1) & $d^{\rm Co-Co}$ = 2.246 & $E$ = -93461.48 & 2.00 \\
(AII)          &           & $d^{\rm Co-C}$ = 1.764        & &      \\
               &   & $h^{\rm Co}$ = 1.200                  & &      \\
               &   & $h^{\rm C}$ = 0.276                   & &      \\
\hline
Co$_2$-SVG     & 1 (P1) & $d^{\rm Co-C}$ = 1.876, 1.876, 1.876 & $E$ = -93462.68 & 1.71  \\
(B)            &   & $h^{\rm Co}$ = 1.171 & &  \\
               &   & $h^{\rm C}$ = 0.0, 0.0, 0.0 & &  \\
\hline
Co$_2$-SVG     & 1 (P1) & $d^{\rm Co-Co}$ = 2.303 & $E$ = -93461.54 & 2.00  \\
(C)            &   & $d^{\rm Co-C}$ = 1.760, 1.760, 1.777 & &  \\
               &   & $h^{\rm Co}$ = 1.238, 2.566 & &  \\
               &   & $h^{\rm C}$ = 0.232, 0.231, 0.586 & &  \\
\hline
Co$_2$-SVG     & 1 (P1) & $d^{\rm Co-Co}$ =2.338 & $E$ = -93461.27 & 1.75  \\
(D)            &   & $d^{\rm Co-C}$ = 1.760, 1.784, 1.763 & &  \\
               &   & $h^{\rm Co}$ = 1.240, 1.986 & &  \\
               &   & $h^{\rm C}$ = 0.154, 0.312, 0.445 & &  \\
\hline
\end{tabular}
\caption{Compilation of geometries, total energies, and spin moments of Co-G,
Co$_2$-G, SVG, Co-SVG, and Co$_2$-SVG structures, obtained in scalar relativistic mode.}
\label{SII}
\end{table}
}
\end{document}